\documentclass[12pt]{article}
\usepackage{amssymb,amscd,amsmath,amsthm}
\usepackage{latexsym,amstext}
\usepackage{latexsym,amstext}
\usepackage{color}
\usepackage{graphicx}
\usepackage{epstopdf}
\usepackage{cite}

\begin{document}

\title{ Redundant poles of the $S$-matrix for the one dimensional Morse potential}

\author{ M. Gadella$^1$, A. Hern\'andez-Ortega$^1$, \c{S}. Kuru$^2$, J. Negro$^1$\\ \\
$^1$Departamento de F\'{\i}sica Te\'orica, At\'omica y Optica  and IMUVA, \\
Universidad de Va\-lladolid, 47011 Valladolid, Spain\\ 
$^2$Department of Physics, Faculty of Science,  \\  Ankara University, 06100
Ankara, Turkey
}

\maketitle

\begin{abstract}

We analyze the structure of the scattering matrix, $S(k)$, for the one dimensional Morse potential. We show that, in addition to a finite number of bound state poles and an infinite number of anti-bound poles, there exist an infinite number of {\it  redundant poles}, on the positive imaginary axis, which do not correspond to either of the other types. This can be solved analytically and exactly. In addition, we obtain wave functions for all these poles and ladder operators connecting them. Wave functions for redundant state poles are connected via two different  series. We also study some exceptional cases.

\end{abstract}






\section{Introduction}
\label{}

As is well known when we deal with non-relativistic quantum scattering, and under some causality conditions \cite{NUS}, the scattering matrix in the momentum representation, $S(k)$, has an analytic continuation to a meromorphic function on the complex plane. Its isolated singularities are poles, which are classified as bound state poles, in one to one correspondence with the bound states, anti-bound poles and resonance poles. However, for some types of potentials other kinds of singularities may arise, like branch cuts and redundant poles. The latter do not correspond to physical states and have been studied already long ago \cite{MA,MA1,MA2}. This work has been continued by some authors and has inspired a bunch of result in scattering theory either for Hermitian or for non-Hermitian Hamiltonians \cite{TER,BIS,VAN,PEI,LR,SIM,MOS}. 

Very recently, Moroz and Miroshnichenko \cite{MM,MM1} had exhaustively studied the analytic behavior of the scattering matrix $S(k)$ corresponding to the radial Schr\"odinger equation with potential

\begin{equation}\label{1}
V(r)=-V_0\,e^{-r/a}\,,
\end{equation}
with $V_0>0$ and $a>0$. The authors find a series of redundant poles of $S(k)$.

These results have in part motivated the discussion presented in this paper. We have started with an exactly solvable one dimensional potential, which is the Morse potential, and study the properties of the scattering matrix, $S(k)$, produced by it. In this case, $S(k)$ can be analytically continued to a meromorphic function on the whole complex plane with an infinite number of simple poles as the only singularities. These poles can be classified into three kinds: i.) A finite number of bound state poles located on the positive imaginary semi axis. ii.) An infinite number of poles on the negative imaginary semi axis, which are usually called the virtual or anti-bound poles.  iii.) Finally, an infinite number of simple poles located along the positive imaginary semi-axis. In exceptional cases, some of these poles may coincide with the bound state poles, as we shall see. No resonance poles are present. 

Then, we may consider the energies $E_i=(\hbar k_i)^2/(2m)$, which correspond to each of the poles $k_i$, and the eigenvalue problem $H\psi_i(x)=E_i\psi_i(x)$. We have obtained the eigenfunctions $\psi_i(x)$ for each of the poles $k_i$. While these eigenfunctions are square integrable for the bound state poles, they are not for all the others. There is also a special point of the spectrum at $E=0$, that some authors call semi-bound state \cite{DOM}.

In previous papers, our group has analyzed ladder operators for exactly solvable models \cite{CEV,GKN,CG}. We do the same here. In general, we obtain three independent series of wave functions related by ladder  operators. One includes wave functions of both bound and antibound poles. For the wave functions for the redundant poles, there exists two independent series and the wave functions inside each of the series are obtained from each other via these ladder operators. We study two exceptional cases, in which bound state and redundant poles may coincide, which produce some unexpected behavior of the ladder operators. 

This article is organized as follows:  In Section 2, we briefly recall the method to solve the one dimensional Schr\"odinger  equation, so as to obtain the wave functions for bound states and scattering states, with stressing on their asymptotic behavior, which is quite relevant for posterior analysis. In section 3, we obtain the scattering matriz $S(k)$ and its singularities.  Section 4 is devoted to the construction and properties of ladder operators. We finish this presentation with some Concluding Remarks. 

\section{The Morse potential}

We begin with a description of the one dimensional Schr\"odinger equation that comes after the Morse potential. This is

\begin{equation}\label{2}
\left[-\frac{d^2}{dx^2} +e^{-2x} -2(A+1/2) e^{-x} \right]\psi(x) =E\psi(x)\,,\quad
A>0, \ \ -\infty <x<\infty \,.
\end{equation}

The exact solvability of this quantum model is based in the fact that the Schr\"odinger equation \eqref{2} may be transformed into confluent hypergeometrical equation after a sequence of changes of variables. 

The result is as follows. Let us make the change of variable as well as
function in the form
\begin{equation}\label{x}
 \psi(x) = e^{-\sqrt{-E} x} e^{-e^{-x}}\, y(z)\,,\quad
 z =2 e^{-x}\,.
\end{equation}
Then, we get
\begin{equation}\label{12}
zy''(z)+(2\sqrt{-E}+1-z)y'(z)-[\sqrt{-E}-A]y(z)=0\,.
\end{equation} 
Now, let us compare \eqref{12} with the standard form of the confluent hypergeometric function, which is
\begin{equation}\label{11}
zy''(z)+(c-z)y'(z)-ay(z)=0\,. 
\end{equation}
We see that these two equations are identical 
with the identification
\begin{equation}\label{13}
c
 =1+2\sqrt{-E}
\,, \qquad a
= -A+\sqrt{-E}\,.
\end{equation}

As is well known, equation \eqref{11} has two linearly independent solutions in terms of the first kind Kummer function:

\begin{equation}\label{14}
y_1(z)=\  _1F_1(a;c;z)\,, \qquad  y_2(z)=z^{1-c}\,_1F_1(a+1-c;2-c;z)\,.
\end{equation}
Note that,
\begin{equation}\label{13}
a':=a+1-c=-A-\sqrt{-E},\qquad c':=2-c=1-2\sqrt{-E}\,.
\end{equation}

We recall that for $c$ integer, we have only one of these solutions and the other is the product of a series times a logarithm. Thus, if $c$ is not an integer, or equivalently, $2\sqrt{-E}$ is not an integer, the general solution of \eqref{2} has the form
\begin{equation}\label{16}
\psi(x) = C_1\,\psi_1(x) + C_2\,\psi_2(x)\,,
\end{equation}
with
\begin{eqnarray}
\psi_1(x)= e^{-\sqrt{-E} x}\, e^{-e^{-x}}\, _1F_1(-A+\sqrt{-E},1+2\sqrt{-E};2e^{-x}) \,,\label{17}\\[2ex]
\psi_2(x)= 2^{-2\sqrt{-E}}\, e^{\sqrt{-E} x}\, e^{-e^{-x}}\,_1F_1(-A-\sqrt{-E},1-2\sqrt{-E};2e^{-x}) \,,\label{18}
\end{eqnarray}
as we may check after a straightforward calculation. 

\begin{figure}
\centering
\includegraphics[width=0.40\textwidth]{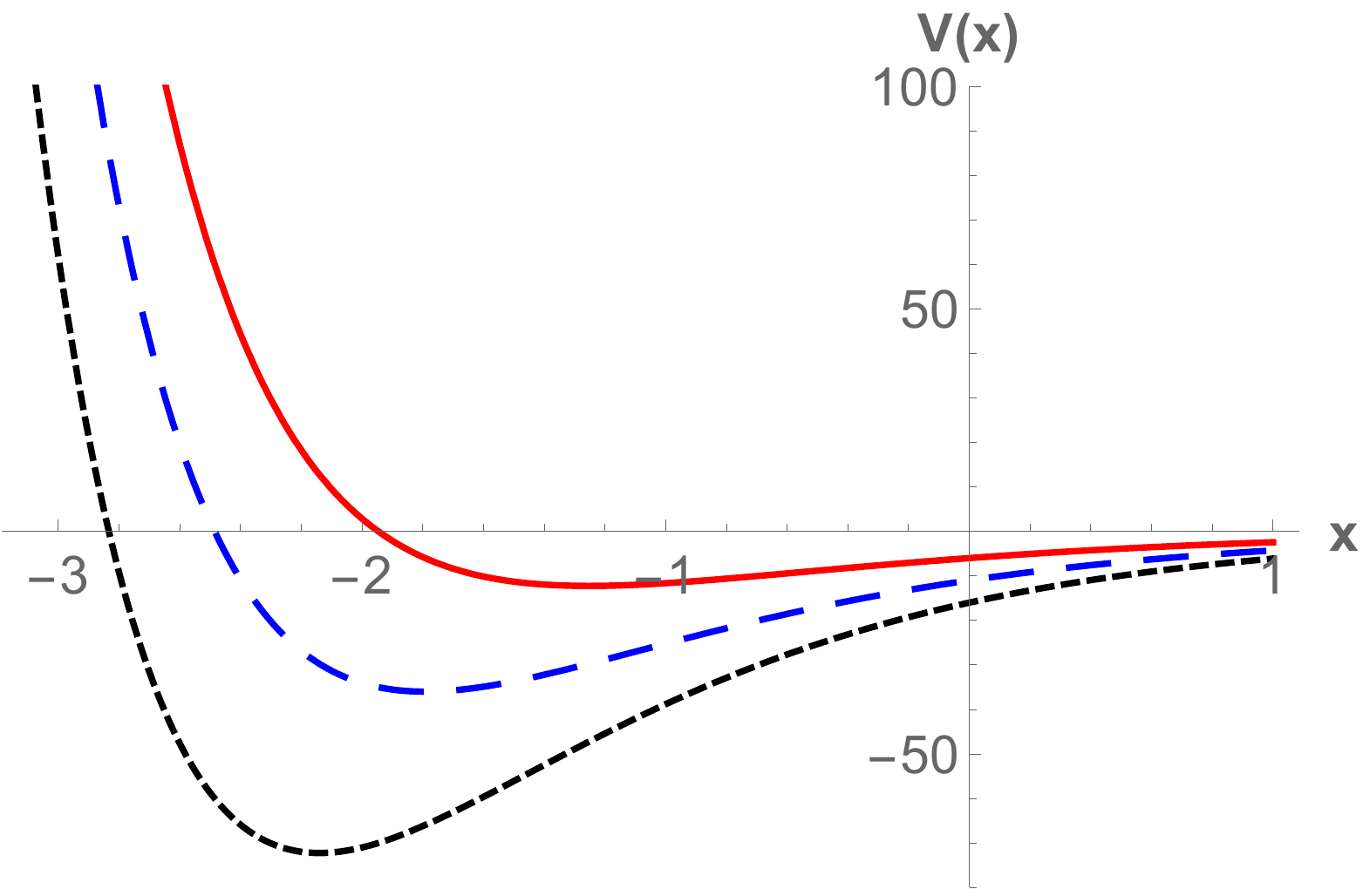}\,\qquad
\caption{\small Plot of the Morse potential for different
values of $A$. The continuous line corresponds to
$A=3$, the dotted line to $A=5.5$, the dashed line to
$A=8$. }\label{pot}
\end{figure}

\subsection{Bound states}

Let us find the solutions of the eigenvalue problem associated to the Morse equation with square integrable eigenfunctions. One way to obtain those solutions is to analyze the behavior of the eigenfunctions in the limits $x\longmapsto \pm \infty$. 

\begin{itemize}

\item{$x \longmapsto +\infty$

In this case, and taking into account that $_1F_1(a,c;0)=1$, the asymptotic behavior of the general solution \eqref{16} is, 

\begin{equation}\label{19}
\psi(x) \approx C_1 e^{-\sqrt{-E}x}+ C_2 \, 2^{-2\sqrt{-E}} e^{\sqrt{-E}x}\,,\qquad x\longmapsto \infty\,.
\end{equation}

Obviously, the second term in the right hand side of \eqref{19} grows exponentially, so that its contribution cannot be square integrable. Consequently, $C_2=0$. The first term decays exponentially provided that $E<0$, which will give a square integrable solution. In consequence, we could keep the solution $\psi(x)=C_1\,\psi_1(x)$ and check its behavior as $x\longmapsto -\infty$.
}

\item{$x \longmapsto -\infty$

We have made the change of variables $z=2e^{-x}$, so that this limit is equivalent to taking the limit $z\longmapsto +\infty$. For large values of $z$, the asymptotic form of $_1F_1(a,b;z)$ is

\begin{equation}\label{20}
_1F_1(a,c;z) \approx \frac{\Gamma(c)}{\Gamma(a)}\,e^z\,z^{a-c}\,, \qquad z\longmapsto +\infty \,,
\end{equation}
so that

\begin{equation}\label{21}
\psi_1(x) \approx \frac{\Gamma(1+2\sqrt{-E})}{\Gamma(-A+\sqrt{-E})}\,2^{-(1+A)-\sqrt{-E}}\,e^{e^{-x}+(1+A)x}\,,\qquad x\longmapsto -\infty\,.
\end{equation}

}

\end{itemize}

The conclusion is clear after \eqref{21}. The solution $\psi_1(x)$ cannot be square integrable unless the coefficient in \eqref{21} vanishes, for which the only possibility is that

\begin{equation}\label{22}
a= -A+\sqrt{-E} =-n,\qquad n=0,1,2\dots\,.
\end{equation}

This gives the following possible values for the energy:

\begin{equation}\label{23}
 E_n= - (A-n)^2, \qquad n=0,1,\dots, N; \quad N=[A]\,,
\end{equation}
where $[A]$ denotes the integer part of $A$ ($[A]<A$). 

The fact that $a$ is now a negative integer shows that the series given the Kummer function truncates and becomes a polynomial. Thus, we have $N$ bound states, which may be expressed in terms of the truncated Kummer function, or alternatively, in terms of the Laguerre associated polynomials as:

\begin{eqnarray}\label{24}
\psi_n(x)= C_{n} e^{-(A-n)x}\, e^{-e^{-x}} {_1F_1(-n;1+2(A-n);2e^{-x})}  \nonumber\\[2ex] = C_n\,e^{-(A-n)x}\, e^{-e^{-x}}\, L_n^{2(A-n)}(2e^{-x}) \,, 
\end{eqnarray}
where $C_n$ are normalization constants. Observe that the eigenvalues $E_n$ must be negative due to our comment above about the limit of the wave function at $+\infty$. Observe that the eigenfunctions $\psi_n(x)$  are products of functions exponentially decreasing as $x\longmapsto \pm\infty$ times a polynomial, hence square integrable. Also note that for bound states $c=1+2(A-n) > 0$. 

\subsection{Scattering wave functions}

The asymptotic analysis of the scattering wave functions may be per se interesting, although our aim lies on its applications in the search of scattering resonances. Let us analyze the asymptotic behavior of solutions of \eqref{2} for positive energy values, $E>0$. If $k=-i\sqrt{-E}$, the explicit form of the two independent solutions of \eqref{2} is given by

\begin{eqnarray}
 \psi_1(x)= e^{-i{k}x}\, e^{-e^{-x}}{_1F_1(-A+ik;1+2ik;2e^{-x})}\,, \label{25} \\[2ex] 
 \psi_2(x)= e^{i{k}x}\, e^{-e^{-x}}2^{-2i{k}}{_1F_1(-A-ik;1-2ik;2e^{-x})}\,. \label{26}
\end{eqnarray}

In the limit $x\longmapsto +\infty$, the Kummer functions, both in \eqref{25} and \eqref{26} go to one, so that the asymptotic form of both solutions for large positive values of $x$ is given by

\begin{equation}\label{27}
\psi_1(x \xrightarrow{}\infty) \approx e^{-ikx} \,, \qquad  \psi_2(x \xrightarrow{} \infty) \approx e^{-i\,2k\log 2} e^{ikx}\,.
\end{equation}

After \eqref{27}, we may consider $\psi_1(x)$ and $\psi_2(x)$ as the incident and reflected wave, respectively. 

In the limit $x\longmapsto -\infty$, we have to take into account the relation \eqref{20}, so as to obtain the following asymptotic forms:

\begin{eqnarray}\label{28}
\psi_1(x\xrightarrow{} -\infty) \approx \frac{\Gamma (1+2ik)}{\Gamma (-A+ik)} 2^{-(1+A)-ik} e^{2e^{-x}+(1+A)x}\,, \nonumber \\[2ex]
\psi_2(x\xrightarrow{} -\infty) \approx \frac{\Gamma (1-2ik)}{\Gamma (-A-ik)} 2^{-(1+A)-ik} e^{2e^{-x}+(1+A)x}\,.
\end{eqnarray}

Observe that, in the limit $x\longmapsto -\infty$, both solutions diverge, although their functional form,  is similar, save for the multiplicative constants. This makes it possible to choose a linear combination of both solutions so that its asymptotic behavior for large negative values equals to zero. This arrives to a relation between the coefficients $C_i$, $i=1,2$ in \eqref{16} as  

\begin{equation}\label{29}
C_2=-\frac{\Gamma(-A-ik)}{\Gamma(-A+ik)}\frac{\Gamma(1+2ik)}{\Gamma(1-2ik)}\,C_{1}\,.
\end{equation}

Thus, we have chosen solutions with the property $\psi(x \xrightarrow{} -\infty)=0$. Due to the form of the Morse potential, this is physically reasonable.

After the above considerations, we may consider $\psi_1(x)$ and $e^{2ik\log 2}\psi_2(x)$ as the incoming and the outgoing wave functions, respectively, which justifies the following the change of notation: $\psi_1(x):= \psi_{\rm IN}(x)$ and $e^{2ik\log 2}\psi_2(x) := \psi_{\rm OUT}(x)$. 

\begin{equation}\label{30}
\psi(x)=C_{\rm IN}\, \psi_{\rm IN}(x)+ C_{\rm OUT}\,\psi_{\rm OUT}(x)\,,
\end{equation}
which after \eqref{29} gives

\begin{equation}\label{31}
  S(k)=-\frac{\Gamma(-A-ik)}{\Gamma(-A+ik)}\frac{\Gamma(1+2ik)}{\Gamma(1-2ik)}\,e^{-2ik \log2}\,.
\end{equation}

Note that for $k$ real, $S(k)$ has modulus one. The phase shift in the momentum representation is given by

\begin{equation}\label{32}
S(k)=e^{i\delta(k)} \Longrightarrow \delta_{A}(k)=-i\log\Big(-\frac{\Gamma(-A-ik)}{\Gamma(-A+ik)}\frac{\Gamma(1+2ik)}{\Gamma(1-2ik)}\Big)-2k\log 2\,.
\end{equation}

In the next Section, we discuss the analytic properties of the function $S(k)$.

\section{The poles of the scattering matrix $S(k)$}

The function $S(k)$ admits an analytic continuation for $k$ complex, which has some singularities, which in our case are simple poles. One of the main objectives of the present paper is the determination and the analysis of properties of these poles, which can be done after equation \eqref{31}. These poles are of four different kinds \cite{NUS,BOH,MM}:

\begin{itemize}

\item{Each of the bound states yields to a simple pole on the positive imaginary semi-axis. In the present model, only a finite number of such poles exist.}

\item{Simple poles on the negative part of the imaginary semi-axis are associated to virtual states, also called antibound states. Only those antibound states, located at values of $ik$ with $k$ near zero, may have observable effects.}

\item{Pairs of poles on the lower half plane, symmetrically located with respect to the negative semi-axis, correspond to scattering resonance states, in other words to quantum unstable states. These poles are often infinite in number and, in principle, they may have arbitrary multiplicity. Poles symmetrically placed with respect to the imaginary axis have the same multiplicity. However, poles with multiplicities higher than two are not computable, in general. Our model does not have resonance poles.}

\item{In addition, poles on the positive imaginary semi-axis, which do not correspond to bound states may exists. These are called {\it redundant poles} of the $S$-matrix. We shall show that the model under study has redundant poles and study some of their properties.}

\end{itemize} 

After the properties of the $\Gamma(z)$ function, we have to look for the singularities in the numerator of \eqref{31}, nevertheless, we have to check whether or not these poles coincide with poles in the denominator, what may lead to regular points. One of the characteristics of the present model is the absence of resonance poles. The absence of resonances is a consequence of the form of the function \eqref{32}, although it  can be surmised after the behavior  of the phase shift $\delta(k)$. Resonances  are usually identified by a sudden and abrupt change of $\delta(k)$ at given energies \cite{BOH}, something that is not observed here, as we may see after Figure 2, in which we have used $\Delta(k)=d(\delta(k))/dk$.

\begin{figure}[htb]
\begin{center}
\includegraphics[width=6cm,height=4.5cm]{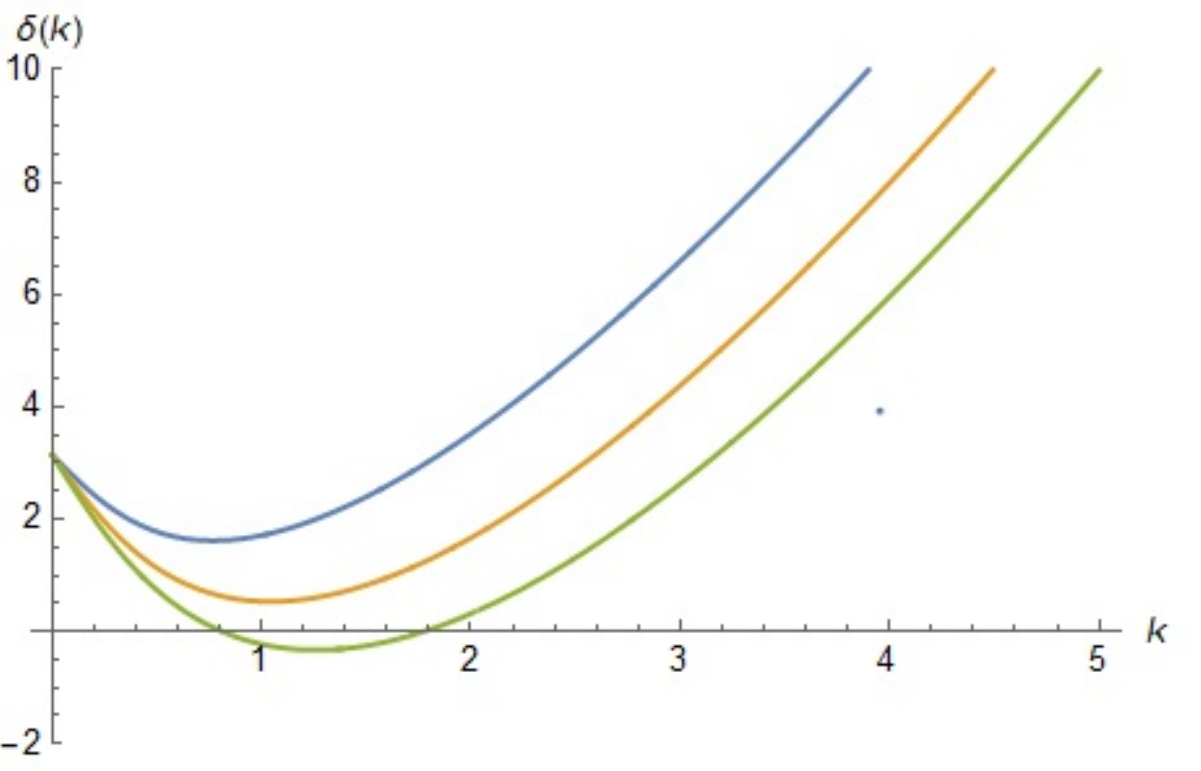}
\ \ \ 
\includegraphics[width=6cm,height=4.5cm]{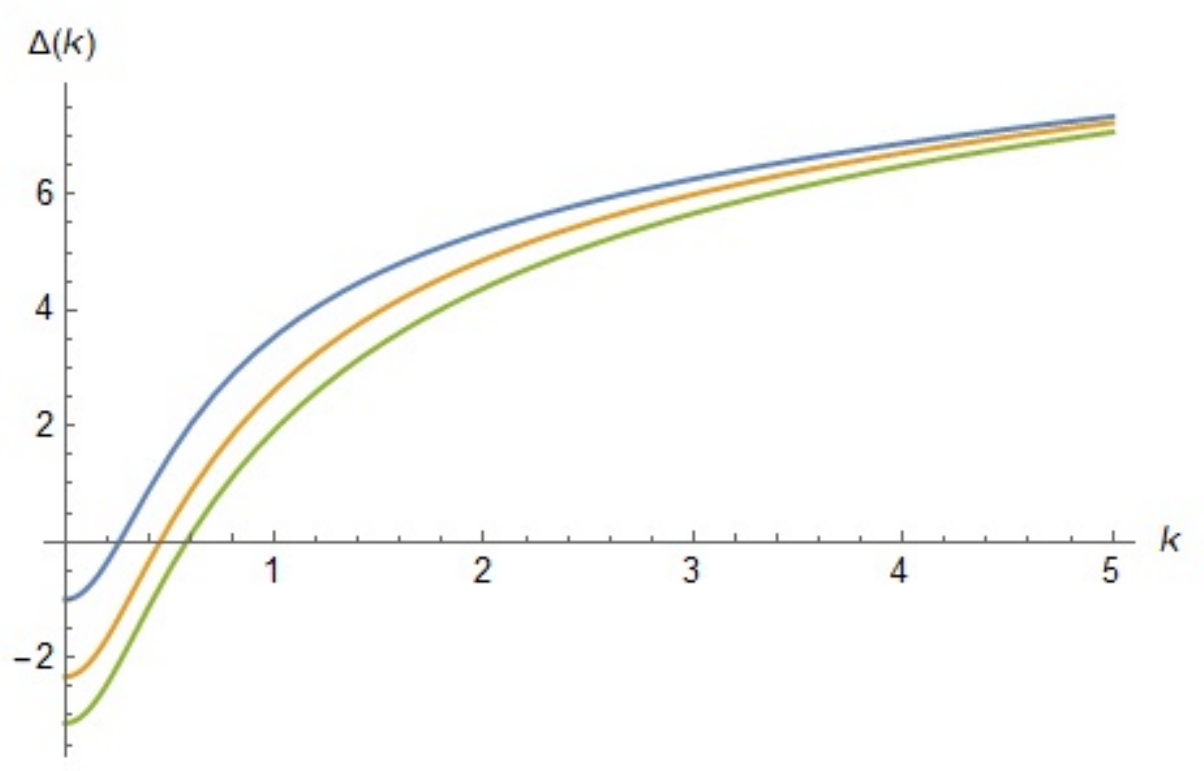}
\caption{\label{Desf2} Phase shifts ($\delta(k)$, left) and its derivative ($\Delta(k)$, right) for different values of $A$: $A=0.5$ (blue), $A=1.5$ (yellow) and $A=2.5$ (Green) .}
\end{center}
\end{figure}

The singularities of $S(k)$ are simple poles on the imaginary axis. Then, we have three possibilities depending on the value of the parameter $A>0$ of the Morse potential, for which we have splited its discussion into the next three Subsections.

\subsection{ General case: $A$ is neither integer nor half-integer}

There are two series of  simple poles that have a different role (see Fig.~\ref{fig3}, left).

\medskip

\noindent
{(a)  First series (or {\bf bound-antibound poles})}. These poles  are located at the points: 
\begin{equation}\label{33}
-A-ik_1 =-n_1 \Longrightarrow k_1= i(A-n_1)\,,\qquad n_{1}=0,1,2,\dots\, .
\end{equation}

About these series of poles, we may say 

\begin{itemize}

\item[(i)] Assume they are of the form \eqref{33} with  $(A-n_1)>0$. 

\noindent
Then, we have a finite number of poles of this type, $k_1(n_1)$, 
for $n_1 =0,\dots, [A]$, on the positive imaginary axis. They are in one to one correspondence with bound states of energies $E(n_1)= (\hbar^2k^2_1(n_1))/(2m)$.

\item[(ii)] The other possibility is that being $k_1$ of the form \eqref{33}, one has that $(A-n_1)<0$. 

Then, there are infinite values of poles, $k_1(n_1)$,
for $n_1 =[A]+1, [A]+2,\dots$, which are located on the negative imaginary axis. These are the so called  anti-bound or virtual state poles.

\end{itemize}

The regular wave function, which correspond to the energies $E(n_1)$, can be trivially obtained from \eqref{26} and takes the form:

\begin{equation}\label{34}
\psi_{n_1}(x) = e^{-(A-n_1)(x-2\log 2)}e^{-e^{-x}}{_1F_1(-n_1;1+2(A-n_1);2e^{-x})}\,.
\end{equation}

\noindent
{(b) Second series: {\bf Redundant poles}}. In addition to the previous series of bound-antibound poles, we have the so called {\it redundant poles}, which are located at the following points on the imaginary axis:

\begin{equation}\label{35}
1+2ik_2 = -n_2  \Longrightarrow k_2 = \frac i2\, (1+n_2)\,,\qquad
n_{2}=0,1,2,\dots\,.
\end{equation}

By reasons which will be clarified soon, we split this series into two sub series as follows:

\begin{itemize}

\item[(i)] Even series: $k_2 = \frac i2\, (1+2 n_2)$, $n_2=0,1,2,\dots$.

\item[(ii)] Odd series: $k_2 =  i (1+ n_2)$, $n_2=0,1,2,\dots$. 

\end{itemize}

Contrarily to the series of bound-antibound poles, the values of the redundant poles are independent of the parameter $A$, as is clear from \eqref{35}. Yet, their properties depend on the form of $A$. For $A$ neither integer nor half integer, there is no overlapping between redundant poles and bound-antibound poles. Then, the regular wave functions for the energies of the redundant poles $E(n_2)$ with $A$ neither integer nor half-integer have the following form:

\begin{equation}\label{36}
\psi_{n_2}(x) = e^{-\frac{n_2+1}2({x}-2\log 2)}e^{-e^{-x}}{_1F_1(-A+\frac{1}{2}(n_2+1);n_2+2;2e^{-x})}\,.
\end{equation}

We discard the second solution, since it has a logarithmic divergence. Next, we analyze the situation for other values of $A$.

\subsection{$A$ is an integer number, $A=N$}

If $A$ is either integer or half-integer, an overlapping between the series of redundant poles and the bound-antibound poles occurs (see Fig.~\ref{fig3}, center).  
Here, it means that a finite number of the poles (a.i) coincide with some redundant poles of the odd list (b.ii). This introduces some modifications on the map of poles, which may be summarized as follows:

\medskip
\noindent
(a) The series of bound-antibound poles is finite. These poles are located at some {\it positive imaginary} points, which are

\begin{equation}\label{37}
-N-ik_1 =-n_1 \Longrightarrow k_1= i(N-n_1)\,,\qquad n_{1}=0,1,2,\dots, N-1\,.
\end{equation}

This has important consequences: Since $N-n_1>0$, we have exactly $N$ bound state poles, or equivalently, $N$ bound states with energies $E(n_1)=(\hbar^2k_1^2(n_1))/(2m)$, $n_{1}=0,1,2,\dots N-1$, and {\it no antibound poles}. In addition, there is a pole with $n_1=N$, which lies at the origin, $k_1=0$, and will be considered later. The regular wave function for the bound states has the form (except for normalization):

\begin{equation}\label{38}
\psi_{n_1}(x) = e^{-(N-n_1)(x-2\log 2)}e^{-e^{-x}}{_1F_1(-n_1;1+2(N-n_1);2e^{-x})}\,,
\end{equation}
for $n_1 =0,\dots, N-1$.  We do not consider the second solution as has a logarithmic divergence. 

\medskip
\noindent
(b) Both sub series of redundant poles have different behavior, i.e., 

\begin{itemize}

\item[(i)] Even series: $k_2 = \frac i2\, (2 n_2+1)$, $n_2=0,1,2,\dots$. 

There is nothing new concerning the even series of redundant poles, which go as in the general case ($A$ neither integer nor half-integer). associated to these poles there is only one regular wave function of hypergeometric type, which is

\begin{equation}\label{39}
\psi_{n_2}(x) = e^{-\frac{2n_2+1}2(x-2\log 2)}e^{-e^{-x}}{_1F_1(-N+\frac{1}{2}(2n_2+1);2n_2+2;2e^{-x})}\,.
\end{equation}

\item[(ii)] Odd series: $k_2 =  i (1+ n_2)$, $n_2=0,1,2,\dots, N-1$.

Here, odd redundant poles exactly match with the bound state poles. Therefore, no odd redundant poles exist in this case.

\end{itemize}

\subsection{$A$ is a half odd number, $A=\frac12(2N-1)$.}

\begin{figure}[h!]
\begin{center}
\includegraphics[width=3cm]{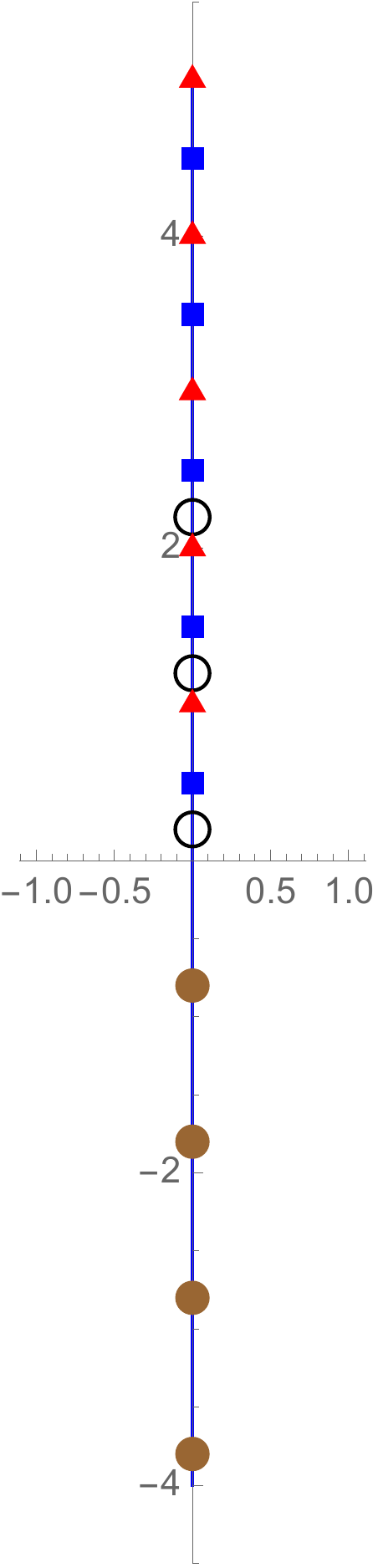}\qquad 
\includegraphics[width=3cm]{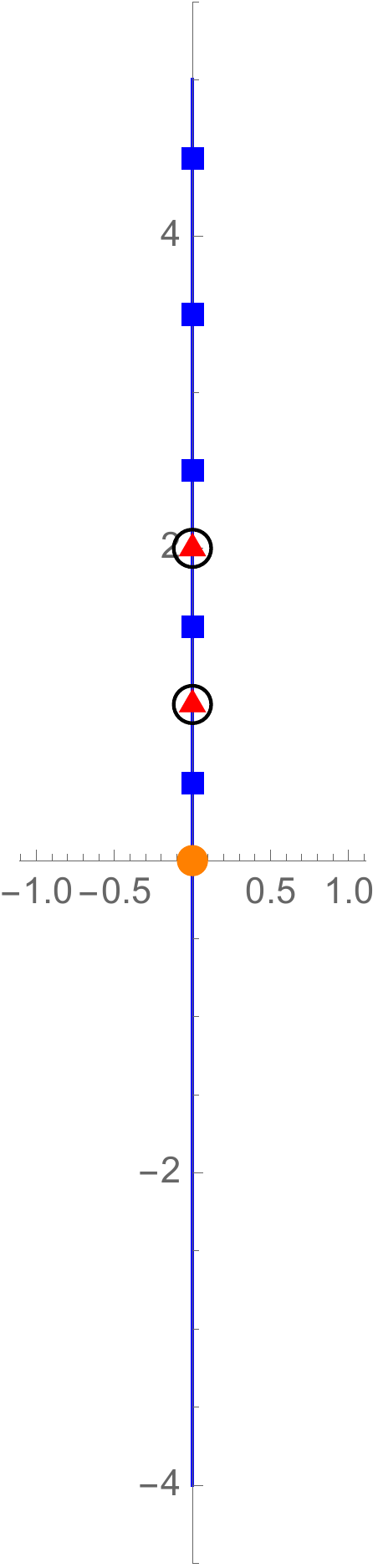}\qquad
\includegraphics[width=3cm]{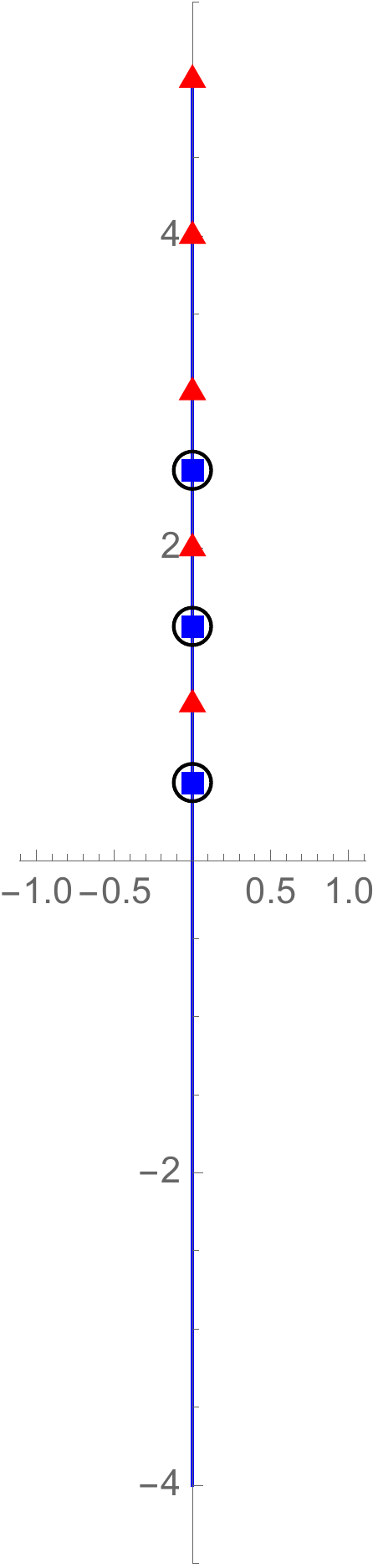}
\caption{\label{fig3} Left: Poles associated to bound (circles), antibound (disks) and redundant states (triangles for even, squares for odd redundant poles) for the general case $A=2.3$. Center: The same for the integer case $A=2$. The orange dot corresponds to the semi--bound state. Right: For the half-integer case $A=2.5$.  The vertical axis is the imaginary part of the complex momentum $k$. }
\end{center}
\end{figure}

Now, the map of poles is similar to the precedent case with $A$ integer
(see Fig.~\ref{fig3}, right), although we may observe some minor still noteworthy differences, as seen from the following analysis:

\medskip
\noindent
(a) First series of bound-antibound poles. 
They are located at the points $k_1= i(\frac 12(2N-1)-n_1)$, with $(\frac 12(2N-1)-n_1)>0$, $n_1 =0,\dots, N-1$. There are bound state poles only and {\it no antibound poles}. The regular wave function for these bound states has the form

\begin{equation}\label{40}
\psi_{n_1}(x) = 
e^{-(N-n_1-1/2)(x-2\log 2)}e^{-e^{-x}}
{_1F_1(-n_1;2(N-n_1);2e^{-x})}\,.
\end{equation}

\medskip\noindent
(b) The two sub series of redundant poles are

\begin{itemize}

\item[(i)] Even series: $k_2 = \frac i2\, (1+2 n_2)$, 
$n_2=0,1,2,\dots, N-1$\,.

The series is finite and redundant poles and bound state poles coincide, so that no even redundant poles really exist.

\item[(ii)] Odd series: $k_2 =  i (n_2+1)$, $n_2=0,1,2,\dots$. This is a sequence of true redundant poles, for which their wave functions take the form:

\begin{equation}\label{41}
\psi_{n_2}(x) = e^{-(n_2+1)(x-2\log 2)}e^{-e^{-x}}
{_1F_1}(-\frac12(2N-1)+(n_2+1);n_2+3;2e^{-x})\,.
\end{equation}

\end{itemize}

\begin{figure}[h!]
\begin{center}
\includegraphics[width=6cm]{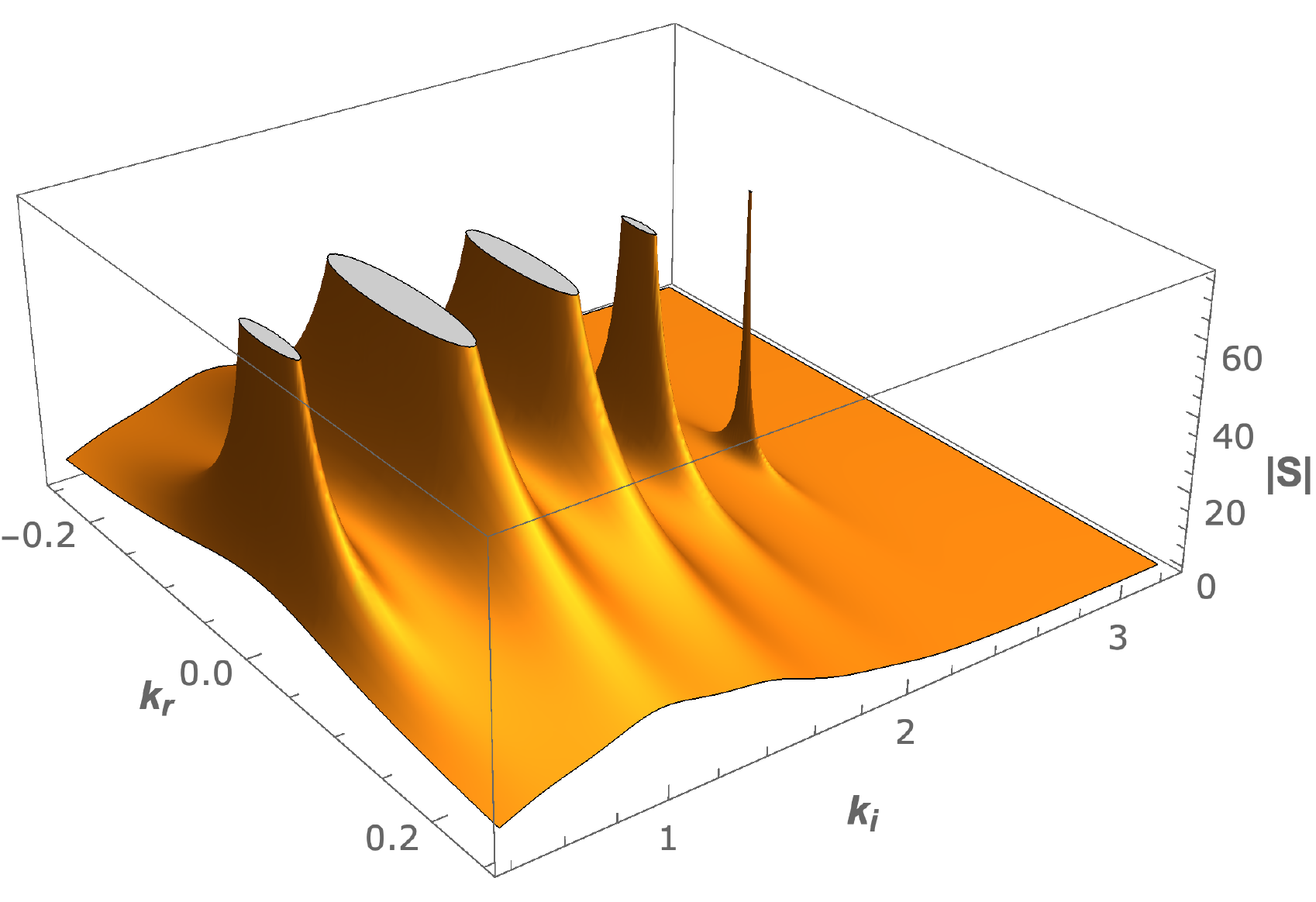}\qquad 
\includegraphics[width=6cm]{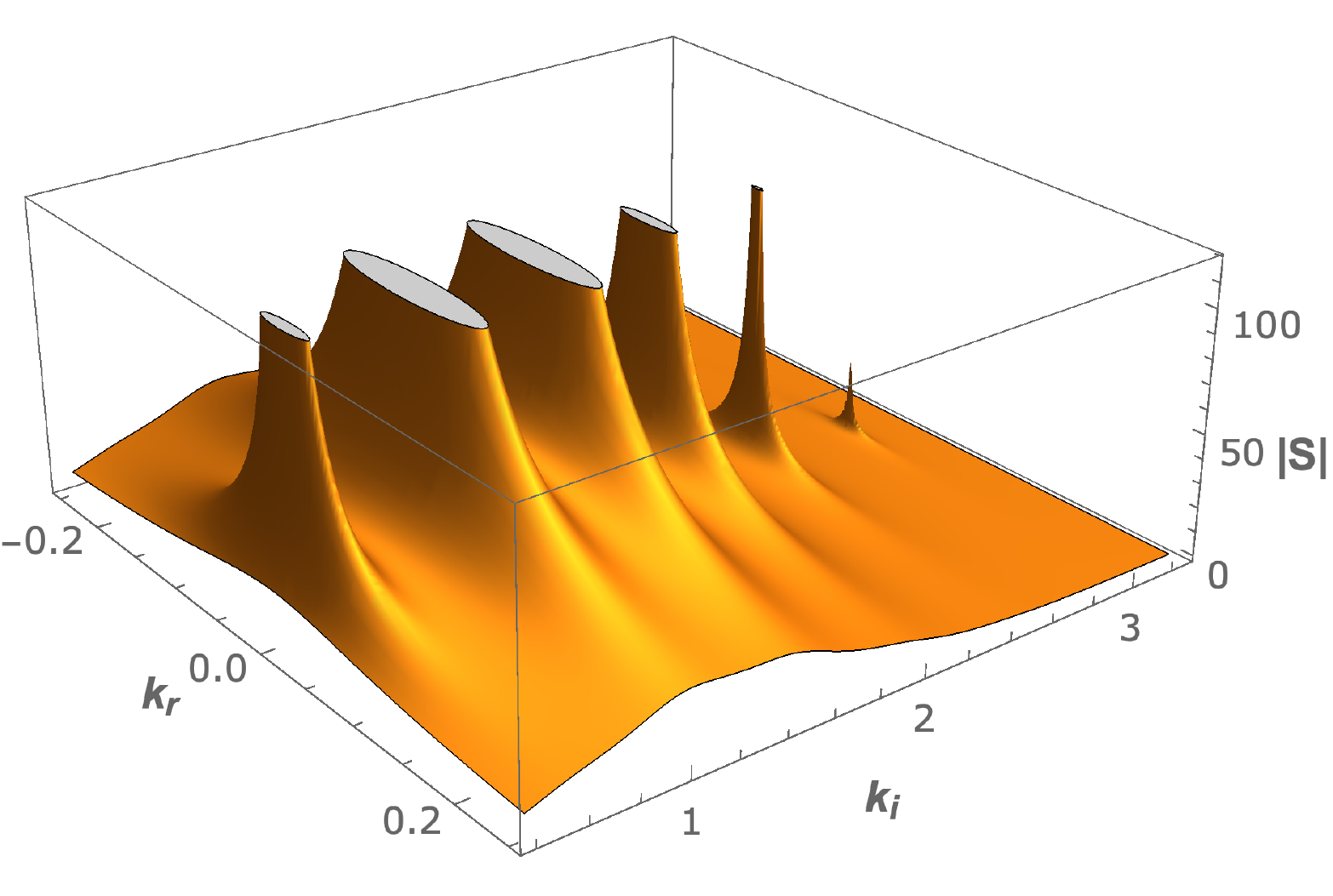}
\caption{\label{fig4} Plot of the absolute value of the $S(k)$ matrix for complex momentum $k=k_r+i\,k_i$:   $A=2$ (left); (b) $A=2.5$ (right). They correspond to the map of poles shown in Fig.~\ref{fig3}, center and right.}
\end{center}
\end{figure}

In Fig.~\ref{fig4} it is shown the poles of the scattering matrix for the
last two special cases: $A$ integer and $A$ half-integer, where bound and some redundant poles coincide.

We conclude at this point the discussion on the bound antibound and redundant poles. It is interesting to remark the existence of wave functions that are the eigenfunctions of the Hamiltonian with eigenvalue equal to the energy $E = (\hbar^2k^2)/(2m)$ where $k$ is the localization of the pole on the complex plane. Only wave functions corresponding to bound state poles are square integrable. In the next Section, we shall construct ladder operators connecting these wave functions. 

\subsection{A remark concerning $E=0$}

Now, we look for solutions to the eigenvalue problem associated to the Schr\"odinger equation \eqref{2} with $E=0$, or equivalently, $k=0$. Obviously, we only need the determination of the eigenfunctions, which must have the form  \eqref{25} or \eqref{26} indistinctly, as both functions coincide as $k=0$. Consequently, it must exist another linearly independent solution for this eigenvalue problem, which comes to be the product of a series times a logarithm. Due to its logarithmic divergency, we discard this second solution, exactly as we did in previous cases. It remains a solution as product of an exponential times a Kummer function, which is

\begin{equation}\label{42}
\psi(x)=  e^{-e^{-x}}{_1F_1(-A;1;2e^{-x})} \,.
\end{equation}

The behavior of the wave function \eqref{40} depends on the values of $A$. If $A$ were a non-negative integer, $A=0,1,2, \dots$, then \eqref{40} goes to zero in the limit $x\longmapsto -\infty$ and is bounded everywhere. Some authors call semi-bounded these type of states \cite{DOM,NEW}. For other values of $A>0$, the function \eqref{40} shows a strong divergence as $x\longmapsto -\infty$, so that it may not have a physical meaning.

\section{Ladder operators}

Let us go back to the Schr\"odinger equation \eqref{2}, in which we restrict ourselves to negative energies $E=-\epsilon^2$. Then, multiply \eqref{2} to the left by $e^{2x}$. After a simple terms rearrangement, \eqref{2} becomes

\begin{equation}\label{43}
h_\epsilon\psi_{\epsilon}(x)\equiv\left[-e^{2x}\frac{d^2}{dx^2} +\epsilon^{2}e^{2x} -2(A+1/2) e^{x} \right]\psi_{\epsilon}(x) =-\psi_{\epsilon}(x)\,,
\end{equation}
where $\psi_\epsilon(x)$ is the eigenfunction of the Hamiltonian $H$ with eigenvalue $-\epsilon^2$ and the meaning of $h_\epsilon$ is obvious. 

Next, let us define two first order differential operators $\mathcal A^+_\epsilon$ and $\mathcal A^-_\epsilon$, having the following form:

\begin{equation}\label{44}
\mathcal A^+_\epsilon := -e^{x}\frac{d}{dx}+\beta e^{x}+\gamma\,,\qquad  \mathcal A^-_\epsilon := e^{x}\frac{d}{dx}+\beta'e^{x}+\gamma'\,,
\end{equation}
such that \eqref{43} becomes 

\begin{equation}\label{45}
h_\epsilon\psi_{\epsilon}(x)=  \left[\mathcal{A}^{+}_{\epsilon}\mathcal{A}^{-}_{\epsilon}+\mathcal{D}_{\epsilon}\right]\psi_{\epsilon}(x)=-\psi_{\epsilon}(x)\,.
\end{equation}

Here, $\beta,\beta',\gamma,\gamma'$ are constants to be determined and $\mathcal{D}_{\epsilon}$ is independent of $x$, although it may depend on $\epsilon$. 
After some straightforward calculations, we obtain 

\begin{eqnarray}\label{46}
\mathcal{A}^{+}_{\epsilon}= -e^{x}\frac{d}{dx}+(1+\epsilon)e^{x}-\frac{1+2A}{2\epsilon+1}\,, \qquad
\mathcal{A}^{-}_{\epsilon}= e^{x}\frac{d}{dx}+\epsilon \, e^{x}-\frac{1+2A}{2\epsilon+1}\,, \nonumber \\[2ex]   \mathcal{D}_{\epsilon}=-\frac{(1+2A)^{2}}{(1+ 2\epsilon)^{2}}\,.
\end{eqnarray}

We may write two different expressions for equation \eqref{43} in terms of the ladder operators \eqref{46}:

\begin{equation}\label{47}
 \left[\mathcal{A}^{+}_{\epsilon}\mathcal{A}^{-}_{\epsilon}+\mathcal{D}_{\epsilon}\right]\psi_\epsilon(x) = \left[\mathcal{A}^{-}_{\epsilon-1}\mathcal{A}^{+}_{\epsilon-1}+\mathcal{D}_{\epsilon-1}\right]\psi_\epsilon(x)=h_\epsilon\,\psi_\epsilon(x) = -\psi_\epsilon(x)\,.
\end{equation}

Then, we can determine a simple intertwining relation between consecutive operators 
$ h_\epsilon$ and $ h_{\epsilon-1}$ through the ladder operators. Let us write

\begin{equation}\label{48}
h_{\epsilon}=\mathcal{A}^{-}_{\epsilon-1}\mathcal{A}^{+}_{\epsilon-1}+\mathcal{D}_{\epsilon-1}\,, \qquad  h_{\epsilon-1}=\mathcal{A}^{+}_{\epsilon-1}\mathcal{A}^{-}_{\epsilon-1}+\mathcal{D}_{\epsilon-1} \,. 
\end{equation}

Then, 

\begin{equation}\label{49}
\mathcal{A}^{+}_{\epsilon-1}h_{\epsilon}=h_{\epsilon-1}\mathcal{A}^{+}_{\epsilon-1}\,, \qquad h_{\epsilon}\,\mathcal{A}^{-}_{\epsilon-1}=\mathcal{A}^{-}_{\epsilon-1}h_{\epsilon-1}\,,
\end{equation}
where the first equation in \eqref{49} is the result of multiplying the first equation in \eqref{48} to the left by $\mathcal A^+_{\epsilon-1}$, while the second equation in \eqref{49} comes from multiplication to the right of \eqref{48} by $\mathcal A^-_{\epsilon-1}$. In both cases, we also have to take into account the second relation in \eqref{48}. 

Relations \eqref{49} have the following important consequence:

\begin{equation}\label{51}
 \mathcal{A}^{+}_{\epsilon-1} \psi_{\epsilon}(x) \propto \psi_{\epsilon-1}\,, \qquad \mathcal{A}^{-}_{\epsilon-1} \psi_{\epsilon-1}(x) \propto \psi_{\epsilon}\,,
\end{equation}
where the symbol $\propto$ means ``equal save for a multiplicative constant'', which is not essential for our purposes \cite{oscar}. 

It is quite important to remark that the signs plus and minus do not correspond necessarily to creation and annihilation operators, respectively. This depends on the relation between $\epsilon$ and the quantum number $n_i$, $i=1,2$. Next, we will study the application of the ladder operators to wave functions for different series of $S(k)$ poles as described in the previous section. 

\subsection{Bound-Antibound series}

Now, in this series and according to (\ref{33}), we have $\epsilon(n_1) =A-n_1$ and $k(n_1) = i\,\epsilon(n_1)$. Since we are discussing each series separately, it is convenient to drop the subindex in $n_{1,2}$ and write  $n$ instead for simplicity. In terms of this index $n$, the operators $\mathcal A^\pm_\epsilon$ are:

\begin{eqnarray}
\mathcal{A}^{+}_{n}= -e^{x}\frac{d}{dx}+(A-n+1)e^{x}-\frac{1+2A}{2(A-n)+1}\,, \label{52}\\[2ex] 
\mathcal{A}^{-}_{n}= e^{x}\frac{d}{dx}+(A-n) e^{x}-\frac{1+2A}{2(A-n)+1}\,, \label{53} \\[2ex]
\mathcal{D}_{n}=-\frac{(1+2A)^{2}}{(1+2(A-n))^{2}}\,. \label{54}
\end{eqnarray}

Since in terms of $n$, $\epsilon(n)=A-n$, we have that 
$\epsilon(n\pm1)=\epsilon(n)\mp 1$ and we conclude that

\begin{equation}\label{54}
\mathcal{A}^{+}_{n+1} \psi_{n}(x) \propto \psi_{n+1}\,, \qquad \mathcal{A}^{-}_{n+1} \psi_{n+1}(x) \propto \psi_{n}\,.
\end{equation}

Therefore, when the operators $\mathcal{A}^{-}_{n}$ and $\mathcal{A}^{+}_{n}$ 
act on bound or antibound states they behave as lowering and raising operators,
respectively, with respect to the index $n$.  Here, three situations are possible:

\begin{itemize}

\item{Regular case. It means that $A$ is neither integer nor semi integer. The ground state wavefunction, $\psi_0(x)$, is annihilated by $\mathcal A_0^-$:  
$\mathcal A^-_0\psi_0(x)=0$. Then, we obtain the wave functions of the $[A]+1$ bound states and afterwards the infinite number of antibound states just by applying the operator $\mathcal A_n^+$, $n=1,2,\dots$ and using the first relation in \eqref{54}. 
}

\item{$A$ is a positive integer, say, $A= N$.  Then, $\epsilon(n) =A-n$ with $n=0,1,\dots, N-1$. The wave functions for bound states, $\psi_0(x), \dots, \psi_{N-1}(x)$, plus the state $\psi_N(x)$, corresponding to the value zero of the energy, are the only wave functions for the bound-antibound series of poles. These wave functions where shown in \eqref{38}. The first bound state also obeys the equation $\mathcal A^-_0\psi_0(x)=0$.

There is another form to write the same set of wave functions by setting $\epsilon(n)=N-n$, this time with $n=N,N+1,\dots,2N$, which gives

\begin{equation}\label{55}
\widetilde \psi_{n}(x) = e^{(N-n)(x-2\log 2)}e^{-e^{-x}}{_1F_1(N-n;1-2(N-n);2e^{-x})}\,.
\end{equation}
It is simple to show that $\psi_{N-n} = \widetilde \psi_{N+n}$, $n=0\dots,N$. Let us fix our attention in the following sequence of repeated wave functions:

\begin{equation}\label{56}
\{\psi_0,\dots,\psi_{N-1}, \psi_N, \widetilde \psi_{N+1},\dots,\widetilde \psi_{2N}\}
\,.
\end{equation}

Let us drop now the tildes in \eqref{56} for simplicity, as we observe that this tilde affects to the wave functions $\psi_n$ with index $n\ge N$. Then,  the ladder operators
$\mathcal A_n^\pm$ act in a natural way on the wave functions of the sequence  \eqref{56}, so as to give:

\begin{equation}\label{57}
\begin{array}{ll}
\mathcal{A}^{+}_{n+1} \psi_{n}(x) \propto \psi_{n+1}\,, \qquad 
&\mathcal{A}^{-}_{n+1} \psi_{n+1}(x) \propto \psi_{n}\,,\qquad
n=0,\dots 2N-1
\\[2.ex]
\mathcal{A}^{+}_{2N+1} \psi_{2N}(x) =0\,, \qquad 
&\mathcal{A}^{-}_{0} \psi_{0}(x) =0\,.
\end{array}
\end{equation}

}

\item{If $A$ is a positive half-integer, the relation between eigenfunctions and ladder operators is similar as in the previous case with $A$ integer. The ladder operators connect the wave functions of the bound states. Note that the wave function for the semi-bound pole at $k=0$ is not included  in the list, as in the previous case for $A=N$ integer. }

\end{itemize}

\subsection{Redundant pole series}

Now, we analyze the action of the ladder operators on the eigenfunctions \eqref{41} with eigenvalues $\epsilon=\frac12(n+1)$, where we have settled $n=n_1$ for simplicity, corresponding to the redundant poles of $S(k)$. Here, we proceed exactly as in the previous cases, replacing the wave functions for the bound-antibound wave function series by the wave functions corresponding to the redundant pole series. Nevertheless, we have one characteristic peculiarity here: ladder operators connect wave functions for even and odd redundant poles separately. Consequently, we analyze each sub-series independently. 

\begin{itemize}

\item{{\it Even series}. We define a new label $m$ so that $n=2m$ and, then, $\epsilon = \frac 12 (n+1)=m+1/2$. In terms of this new label $m$, we have the following explicit expressions for the ladder operators:

\begin{eqnarray}\label{58}
\mathcal{A}^{+}_{m}= -e^{x}\frac{d}{dx}+(m+\frac{3}{2})e^{x}-\frac{1+2A}{2(m+1)}\,, \nonumber\\[2ex]  \mathcal{A}^{-}_{m}= e^{x}\frac{d}{dx}+(m+\frac{1}{2}) e^{x}-\frac{1+2A}{2(m+1)} \,, \nonumber 
\\[2ex] \mathcal{D}_{m}=-\frac{(1+2A)^{2}}{4(m+1)^{2}}\,.
\end{eqnarray}

The wave functions corresponding to these even redundant poles are given by

\begin{equation}\label{59}
\psi_m(x) = e^{(2m+1)(Ln(2)
    -\frac{x}{2})}e^{-e^{-x}}{_1F_1(-A+(m+1/2);2m+2;2e^{-x})}\,,\;
    m\geq0\,.
\end{equation}

As we have done for the bound-antibound series, we may consider {\it tilded} wave functions labelled with negative values of $m$ and defined for $m\le -1$ as

\begin{equation}\label{60}
\widetilde\psi_m(x) = e^{-(2m+1)(Ln(2)
    -\frac{x}{2})}e^{-e^{-x}}{_1F_1(-A-(m+1/2);-2m;2e^{-x})}\,. 
\end{equation}

It is easy to check that

\begin{equation}\label{61}
\tilde \psi_m= \tilde \psi_{-m-1}, \qquad n=0,1,2,\dots\,,
\end{equation}
expressions that define a double sequence of wave functions given by

\begin{equation}\label{62}
\{ \dots,\psi_1,\psi_0,\tilde \psi_{-1}, \tilde \psi_{-2},\dots \}\,.
\end{equation}

As before, we drop the tildes in \eqref{62}, so as to write the action of the ladder operators in a compact form, which should not be cause for confusion. This action can be written as

\begin{equation}\label{63}
\mathcal{A}^{+}_{n-1} \psi_{n}(x) \propto \psi_{n-1}\,, \qquad 
\mathcal{A}^{-}_{n-1} \psi_{n-1}(x) \propto \psi_{n}\,,
\qquad n\in \mathbb Z \,,
\end{equation}
where $\mathbb Z$ is the set of integer numbers, either positive or negative. here one should have taken into account that $\epsilon(m\pm 1) = \epsilon(m)\pm 1$. 

Also observe another peculiarity for these series: {\it There exists no fundamental or ground state state to be annihilated by any of the ladder operators}. 
 
}

\item{

{\it Odd series}. Now, the new label $m$ should be given by $n=2m+1$, so that $\epsilon =\frac12(n+1)=m+1$. For completeness, we include the explicit form of the ladder operators in terms of $m$ for the odd series:

\begin{eqnarray}\label{64}
\mathcal{A}^{+}_{m}= -e^{x}\frac{d}{dx}+(m+2)e^{x}-\frac{1+2A}{2m+3}\,,\nonumber\\[2ex]   \mathcal{A}^{-}_{m}= e^{x}\frac{d}{dx}+(m+1) e^{x}-\frac{1+2A}{2m+3}  \,, \nonumber\\[2ex]   \mathcal{D}_{m}=-\frac{(1+2A)^{2}}{(2m+3)^{2}}\,.
\end{eqnarray}

The non-normalized wave function labelled by the index $m$ is now,

\begin{equation}\label{63}
\psi_m(x) = e^{(2m+2)(Ln(2)-\frac{x}{2})}e^{-e^{-x}}{_1F_1(-A+m+1;2m+3;2e^{-x})}\,. 
\end{equation}

The remainder of the discussion goes exactly as in the even series, so that we omit it. }

\end{itemize}

We have completed the discussion on the ladder operators for all series of eigenwave functions.

\section{Concluding remarks}

We have studied the analytic properties of the scattering matrix $S(k)$ for the one dimensional Morse potential, which is one on a list of one dimensional potentials \cite{COO} producing an exactly solvable Schr\"odinger equation. We have shown that all poles are simple and lie on the imaginary axis, which implies the absence of resonances. 

All poles of the scattering matrix are simple and their classification and behavior depends on the form of the parameter $A$ of the potential, see equation \eqref{2}. In terms of this classification, there are three groups that we consider separately: Either $A$ is integer, or half integer or neither of both. When $A$ is neither integer nor half integer, we have obtained two independent series of poles, the former is the set of bound-antibound poles. Bound poles lie on the positive imaginary semi-axis and are in one to one correspondence with bound states, while antibound poles lie on the negative imaginary axis and are infinite in number. The second one, is the list of {\it redundant poles}, located along the positive imaginary axis. For the values of $A$ considered, redundant poles are independent of bound state poles and the only information they bring us is the absence of a basis of solutions made of hypergeometric functions. See illustration in Figure 4. 

When $A$ is an integer or a half integer number, we have observed an anomalous behavior in the sense that some of the redundant poles and the bound state poles overlap. This is a much more interesting situation. This coincidence is a kind of interference that eliminates an infinite set of the redundant sub-series (where the coincidence applies), as well as the infinite set of antibound states. The finite number of bound poles are at the same time redundant, i.e., the second independent solution is not hypergeometric.

All these poles are located at imaginary momentum values $ik$ with energies $E=(\hbar^2 k^2)/(2m)$. These energies may be looked as eigenvalues of the Hamiltonian and we obtain their corresponding eigenfunctions as solutions of the Schr\"odinger equation. Since the Schr\"odinger equation is of order two, it has two independent solutions: one is the product of an exponential times a Kummer function and the other has a logarithmic divergence, so that we have discarded the latter in all cases. Only those eigenfunctions in correspondence with bound state poles are square integrable.

Following previous works by our group \cite{KN,CEV,GKN}, we have constructed ladder operators that connect the eigenfunctions in correspondence with the $S(k)$ poles. We have shown that, in the most general case, there exists three different chains of eigenfunctions connected by the ladder operators: the eigenfunctions for the bound-antibound poles belong to the first chain. Here, there is a ground state, which plays the same role as the ground state in the harmonic oscillator.   We also analyze the exceptional cases in which $A$ is either integer or half integer, where the results are similar with some variations. In these excepcional cases the ladder operators connect the finite sequence of bound states in a double way. This behaviour remind us of the $su(2)$ Lie algebra corresponding to integer ($A$ integer) or half-integer ($A$ half-integer) spin.

Then, there are two other chains of eigenfunctions of redundant poles, depending on the even or odd character of a label, which are mutually independent. Here there are some peculiarities, such that the {\it  non-existence of a ground state}. Creation and annihilation connect doubly infinite series of eigenfunctions. 

 \section*{Acknowledgments}

This work was partially supported by Junta de Castilla y Le\'on (BU229P18, VA137G18).  \c{S}.~Kuru acknowledges Ankara University and the warm hospitality at Department
of Theoretical Physics, University of Valladolid, where this work has been done.





\end{document}